\documentstyle[11pt]{article}

\newtheorem{theorem}{Theorem}[section]
\newtheorem{lemma}{Lemma}[section]
\newtheorem{corollary}{Corollary}[section]
\newcommand{\cents}{\kappa}

\def \lket {\left|}
\def \rket {\right\rangle}
\newcommand{\ket}[1]{\lket #1\rket}
\newcommand{\comment}[1]{}
\def \Qspace {l_2(Q)}
\def\qed{$\Box$}

\begin{document}

\title{\bf
Probabilities to accept languages
by quantum finite automata}

\author{
Andris Ambainis\thanks{ Supported
by Berkeley Fellowship for Graduate Studies.}\\
Computer Science Division\\
University of California\\
Berkeley, CA 94720-2320\\
e-mail:{\tt ambainis@cs.berkeley.edu}
\and 
Richard Bonner\\
Department of Mathematics and Physics\\
M\" alardalens University, Sweden\\
e-mail:{\tt richard.bonner@mdh.se}
\and
R\=usi\c n\v s Freivalds\thanks{
 Research supported by Grant No.96.0282 from the
 Latvian Council of Science}\\
Institute of Mathematics and Computer Science\\
University of Latvia\\
Rai\c na bulv. 29, Riga, Latvia\\
e-mail:{\tt rusins@cclu.lv} 
\and 
Arnolds \c Kikusts\thanks{
 Research supported by Grant No.96.0282 from the
 Latvian Council of Science}\\
Institute of Mathematics and Computer Science\\
University of Latvia\\
Rai\c na bulv. 29, Riga, Latvia\\
e-mail:{\tt sd70053@lanet.lv} }

\date{}
\maketitle

\bibliographystyle{alpha}

\begin{abstract}
We construct a hierarchy of regular languages such that the current
language in the hierarchy can be accepted by 1-way quantum finite automata
with a probability smaller than the corresponding probability for the
preceding language in the hierarchy. These probabilities converge to
$\frac{1}{2}$.

\end{abstract}

\section{Introduction}

Quantum computation is a most challenging project involving research both
by physicists and computer scientists. The principles of quantum
computation differ from the principles of classical computation very much.
The classical computation is based on classical mechanics while
quantum computation attempts to exploit phenomena specific to
quantum physics.

One of features of quantum mechanics is that a quantum process can be
in a combination (called {\em superposition}) of several states and these
several states can interact one with another.
A computer scientist would call this {\em a massive
parallelism}.
This possibility of massive parallelism is very important for
Computer Science. In 1982, Nobel prize winner physicist Richard Feynman
(1918-1988) asked what effects the principles of quantum
mechanics can have on computation\cite{Fe 82}. An exact simulation of quantum
processes demands exponential running time. 
Therefore, there may be other computations  
which are performed nowadays by classical computers but might be
simulated by quantum processes in much less time.

R.Feynman's influence was (and is) so high that rather soon this
possibility was explored both theoretically and practically. 
David Deutsch\cite{De 89} introduced quantum Turing machines,
quantum physical counterparts of probabilistic Turing machines.
He conjectured that they may be more efficient that classical 
Turing machines. He also showed  the existence of a universal quantum 
Turing machine. This construction was subsequently improved 
by Bernstein and Vazirani \cite{BV 97} and Yao \cite{Ya 93}.  

Quantum Turing machines might have remained relatively unknown
but two events caused a drastical change.
First, Peter Shor \cite{Sh 94} invented surprising polynomial-time
quantum algorithms for computation of discrete logarithms and for
factorization of integers.
Second, 
unusual quantum circuits having no
classical counterparts (such as quantum bit teleportation) have
been physically implemented. Hence, there is a chance that 
universal quantum computers may be built. 
Moreover, since the modern public-key
cryptography is based on intractability of discrete logarithms and
factorization of integers, building a quantum computer implies
building a code-breaking machine.

In this paper, we consider quantum finite automata (QFAs),
a different model of quantum computation.
This is a simpler model than quantum Turing machines and
and it may be simpler to implement.

Quantum finite automata have been studied in \cite{AF 98,BP 99,KW 97,MC 97}. 
Surprisingly, QFAs do not generalize deterministic finite automata. 
Their capabilities are incomparable.
QFAs can be exponentially more space-efficient\cite{AF 98}.
However, there are regular languages that cannot be recognized by quantum
finite automata\cite{KW 97}.

This weakness is caused by reversibility.
Any quantum computation is performed by means of unitary
operators. One of the simplest properties of these operators
shows that such a computation is reversible. The result always determines
the input uniquely. It may seem to be a very strong limitation. 
Luckily, for unrestricted quantum algorithms (for instance,
for quantum Turing machines) this is not so.
It is possible to embed any irreversible computation in an appropriate  
environment which makes it reversible\cite{Be 89}. 
For instance, the computing agent
could keep the inputs of previous calculations in successive order.
Quantum finite automata are more sensitive to the reversibility requirement.

If the probability with which a QFA is required to be correct
decreases, the set of languages that can be recognized increases.
In particular\cite{AF 98}, there are languages that can be recognized 
with probability 0.68 but not with probability 7/9.
In this paper, we extend this result by constructing a hierarchy of
languages in which each next language can be recognized with
a smaller probability than the previous one.

\section{Preliminaries}

\subsection{Basics of quantum computation}
\label{basics}

To explain the difference between classical and quantum mechanical
world, we first consider one-bit systems.
A classical bit 
is in one of two
classical states $true$ and $false$.
A {\em probabilistic} counterpart of the
classical bit can be $true$ with a
probability $\alpha $ and $false$ with probability $\beta $, where
$\alpha + \beta = 1$. 
A {\em quantum bit (qubit)} is very much like to it
with the following distinction. 
For a {\em qubit} $\alpha $ and $\beta $ can be 
arbitrary complex numbers with the property
$\|\alpha \|^2 + \|\beta \|^2 = 1$.
If we observe a qubit, we get $true$ with probability
$\|\alpha\|^2$ and $false$ with probability $\|\beta\|^2$, just like
in probabilistic case. 
However, if we modify a quantum system without observing it
(we will explain what this means), the set of transformations
that one can perform is larger than in the probabilistic case.
This is where the power of quantum computation comes from.

More generally, we consider quantum systems with $m$ basis states.
We denote the basis states $\ket{q_1}$, $\ket{q_2}$, $\ldots$,
$\ket{q_m}$. Let $\psi$ be a linear combination of them
with complex coefficients
\[ \psi=\alpha_1\ket{q_1}+\alpha_2\ket{q_2}+\ldots+\alpha_m\ket{q_m} .\]
 
The $l_2$ norm of $\psi$ is 
\[  \|\psi\|=\sqrt{|\alpha_1|^2+|\alpha_2|^2+\ldots+|\alpha_m|^2}. \]

The state of a quantum system can be any $\psi$ with $\|\psi\|=1$.
$\psi$ is called a {\em superposition} of $\ket{q_1}$, $\ldots$, $\ket{q_m}$.
$\alpha_1$, $\ldots$, $\alpha_m$ are called {\em amplitudes} of
$\ket{q_1}$, $\ldots$, $\ket{q_m}$.
We use $\Qspace$ to denote the vector space consisting 
of all linear combinations of $\ket{q_1}$, $\ldots$, $\ket{q_m}$.

Allowing arbitrary complex amplitudes is essential for physics.
However, it is not important for quantum computation.
Anything that can be computed with complex amplitudes can be done
with only real amplitudes as well.
This was shown for quantum Turing machines in \cite{BV 93}\footnote{For
unknown reason, this proof does not appear in \cite{BV 97}.}
and the same proof works for QFAs.  
However, it is important that {\em negative} amplitudes are allowed.
For this reason, we assume that
all amplitudes are (possibly negative) reals.

There are two types of transformations that can be performed
on a quantum system. The first type are unitary transformations.
A unitary transformation is a linear transformation $U$ on $\Qspace$
that preserves $l_2$ norm. (This means that any $\psi$ with $\|\psi\|=1$
is mapped to $\psi'$ with $\|\psi'\|=1$.)

Second, there are measurements. 
The simplest measurement is observing 
$\psi=\alpha_1\ket{q_1}+\alpha_2\ket{q_2}+\ldots+\alpha_m\ket{q_m}$
in the basis $\ket{q_1}, \ldots, \ket{q_m}$.
It gives $\ket{q_i}$ with probability $\alpha_i^2$.
($\|\psi\|=1$ guarantees that probabilities of different outcomes sum to 1.)
After the measurement, the state of the system changes to $\ket{q_i}$
and repeating the measurement gives the same state $\ket{q_i}$.

In this paper, we also use {\em partial measurements.}
Let $Q_1, \ldots, Q_k$ be pairwise disjoint subsets of $Q$ such that
$Q_1\cup Q_2 \cup \ldots \cup Q_k=Q$.
Let $E_j$, for $j\in\{1, \ldots, k\}$, denote the 
subspace of $\Qspace$ spanned by $\ket{q_j}$, $j\in Q_i$.
Then, a {\em partial measurement} w.r.t. $E_1, \ldots, E_k$
gives the answer $\psi\in E_j$ with probability $\sum_{i\in Q_j} \alpha_i^2$.
After that, the state of the system collapses to the projection
of $\psi$ to $E_j$.
This projection is $\psi_j= \sum_{i\in Q_j}  \alpha_i \ket{q_i}$.

\subsection{Quantum finite automata}

Quantum finite automata were introduced twice. First this was done by C.
Moore and J.P.Crutchfield \cite{MC 97}. Later in a different and
non-equivalent way these automata were introduced by A. Kondacs and J.
Watrous \cite{KW 97}.

The first definition just mimics the definition of
1-way probabilistic finite automata 
only substituting {\em stochastic} matrices by {\em unitary} ones. 
We use a more elaborated definition  \cite{KW 97}.

A QFA is a tuple
$M=(Q;\Sigma ;V ;q_{0};Q_{acc};Q_{rej})$ where $Q$ is a finite set
of states, $\Sigma $ is an input alphabet, $V$ is a transition function,
$q_{0}\in Q$ is a starting state, and $Q_{acc}\subset Q$
and $Q_{rej}\subset Q$ are sets of accepting and rejecting states.
The states in $Q_{acc}$ and $Q_{rej}$ are called {\em halting states} and
the states in $Q_{non}=Q-(Q_{acc}\cup Q_{rej})$ are called
{\em non halting states}.
$\kappa$ and $\$$ are symbols that do not belong to $\Sigma$.
We use $\kappa$ and $\$$ as the left and the right endmarker,
respectively. The {\em working alphabet} of
$M$ is $\Gamma = \Sigma \cup \{\kappa ;\$\}$.


The transition function $V$ is a mapping from $\Gamma\times \Qspace$
to $\Qspace$ such that, for every $a\in\Gamma$, the function
$V_a:\Qspace\rightarrow\Qspace$ defined by $V_a(x)=V(a, x)$ is a 
unitary transformation.

The computation of a QFA starts in the superposition $|q_{0}\rangle$.
Then transformations corresponding to the left endmarker $\kappa$,
the letters of the input word $x$ and the right endmarker $\$$ are
applied. The transformation corresponding to $a\in \Gamma$ consists
of two steps.

1. First, $V_{a}$ is applied. The new superposition $\psi^{\prime}$
is $V_{a}(\psi)$ where $\psi$ is the superposition before this step.

2. Then, $\psi^{\prime}$ is observed with respect to 
$E_{acc}, E_{rej}, E_{non}$ where
$E_{acc}=span\{|q\rangle:q\in Q_{acc}\}$,
$E_{rej}=span\{|q\rangle :q\in Q_{rej}\}$,
$E_{non}=span\{|q\rangle :q\in Q_{non}\}$
(see section \ref{basics}).

If we get $\psi^{\prime} \in E_{acc}$, the input is accepted.
If we get $\psi^{\prime} \in E_{rej}$, the input is rejected.
If we get $\psi^{\prime} \in E_{non}$, the next transformation is applied.

We regard these two transformations as reading a letter $a$.
We use $V'_a$ to denote the transformation consisting of
$V_a$ followed by projection to $E_{non}$. 
This is the transformation mapping $\psi$ to the non-halting part
of $V_a(\psi)$. 
We use $\psi_y$ to denote the non-halting part of 
QFA's state after reading the left endmarker $\kappa$ and the
word $y\in\Sigma^*$.

\comment{For probabilistic computation, the probability
of correct answer can be easily increased 
by executing several copies of an automaton (or Turing machine)
in parallel. Hence, if a language can be recognized by a probabilistic

Hence, it is not surprising that \cite{KW 97} wrote
"with error probability bounded away from $1/2$", thinking
that all such probabilities are equivalent.
However, mixing reversible (quantum computation) and non-reversible
(measurements after each step) components in one model
makes it impossible.}
We compare QFAs with different probabilities of correct answer.
This problem was first considered by A. Ambainis
and R. Freivalds\cite{AF 98}. The following theorems were proved there:

\begin{theorem}
\label{T1}
Let $L$ be a language and $M$ be its minimal automaton.
Assume that there is a word $x$ such that $M$  
contains states $q_1$, $q_2$ satisfying:
\begin{enumerate}
\item
$q_1\neq q_2$,
\item
If $M$ starts in the state $q_1$ and reads $x$,
it passes to $q_2$,
\item
If $M$ starts in the state $q_2$ and reads $x$,
it passes to $q_2$, and
\item
$q_2$ is neither "all-accepting" state, nor "all-rejecting" state.
\end{enumerate}
Then $L$ cannot be recognized by a 1-way quantum finite automaton with
probability $7/9+\epsilon$ for any fixed $\epsilon>0$.
\end{theorem}

\begin{theorem}
\label{T2}
Let $L$ be a language and $M$ be its minimal automaton.
If there is no $q_1, q_2, x$ satisfying conditions of Theorem \ref{T1}
then $L$ can be recognized by a
1-way reversible finite automaton (i.e. $L$ can be recognized by a 1-way
quantum finite automaton with probability 1).
\end{theorem}

\begin{theorem}
\label{T4}
The language $a^{*}b^{*}$ can be recognized by a 1-way QFA
with the probability of correct answer $p=0.68...$ where
$p$ is the root of $p^3+p=1$.
\end{theorem}

\begin{corollary}
\label{C1}  
There is a language that can be recognized by a 1-QFA with probability
$0.68...$ but not with probability $7/9+\epsilon$.
\end{corollary}

For probabilistic automata, the probability of correct answer can be 
increased arbitrarily and this property of probabilistic computation
is considered as evident. Theorems above show thatits counterpart
is not true in the quantum world! The reason for that is that the model
of QFAs mixes reversible (quantum computation) components with
nonreversible (measurements after every step).

In this paper, we consider the best probabilities of acceptance by 1-way
quantum finite  automata the languages $a^{*}b^{*}\dots z^{*}$. Since the
reason why the language $a^{*}b^{*}$ cannot be accepted by 1-way 
quantum finite  automata is the property described in the Theorems
\ref{T1} and \ref{T2}, this new result provides an insight on what the
hierarchy
of languages with respect to the probabilities of their acceptance by
1-way quantum finite automata may be. 
We also show a generalization of Theorem \ref{T4} in a style similar
to Theorem \ref{T2}.

\section{Main results}

\begin{lemma}
\label{lemma}
For arbitrary real $x_1>0$, $x_2>0$, ..., $x_n>0$, there exists
a unitary $n\times n$ matrix $M_n(x_1,x_2,...,x_n)$ with elements $m_{ij}$ 
such that
$$
m_{11}=\frac{x_1}{\sqrt{x_1^2+...+x_n^2}},\
m_{21}=\frac{x_2}{\sqrt{x_1^2+...+x_n^2}},\ ...,
m_{n1}=\frac{x_n}{\sqrt{x_1^2+...+x_n^2}}.
$$
\end{lemma}
\hspace*{118mm} \qed

Let $L_n$ be the language $a_1^*a_2^*...a_n^*$.
\begin{theorem}
\label{part1}
The language $L_n$ ($n>1$) can be recognized by a 1-way QFA
with the probability of correct answer $p$ where $p$ is the root of
$p^\frac{n+1}{n-1}+p=1$ in the interval $[1/2, 1]$.
\end{theorem}
{\bf Proof:}
Let $m_{ij}$ be the elements of the matrix $M_k(x_1,x_2,...,x_k)$
from Lemma \ref{lemma}.
We construct a $k\times (k-1)$ matrix $T_k(x_1,x_2,...,x_k)$
with elements $t_{ij}=m_{i,j+1}$.
Let $R_k(x_1,x_2,...,x_k)$ be a $k\times k$ matrix
with elements $r_{ij}=\frac{x_i\cdot x_j}{x_1^2+...+x_k^2}$
and $I_k$ be the $k \times k$ identity matrix.

For fixed $n$,
let $p_n\in[1/2, 1]$ satisfy $p_n^\frac{n+1}{n-1}+p_n=1$
and $p_k$ ($1\leq k<n$) = $p_n^\frac{k-1}{n-1}-p_n^\frac{k}{n-1}.$
It is easy to see that $p_1+p_2+...+p_n=1$ and
\begin{equation}
1-\frac{p_n(p_k+...+p_n)^2}{(p_{k-1}+...+p_n)^2}=
1-\frac{p_np_n^\frac{2(k-1)}{n-1}}{p_n^\frac{2(k-2)}{n-1}}=
1-p_n^\frac{n+1}{n-1}=p_n.
\end{equation}

Now we describe a 1-way QFA accepting the language $L_n$.

The automaton has $2n$ states:
$q_1$, $q_2$, ... $q_{n}$ are non halting states,
$q_{n+1}$, $q_{n+2}$, ... $q_{2n-1}$ are rejecting states
and $q_{2n}$ is an accepting state.
The transition function is defined by unitary block matrices
$$V_\kappa=
\left (
\begin{array}{cc}
M_n(\sqrt{p_1},\sqrt{p_2},...,\sqrt{p_n})&{\bf 0}\\
{\bf 0}&I_n
\end{array}
\right ),
$$
$$V_{a_1}=
\left (
\begin{array}{ccc}
R_n(\sqrt{p_1},\sqrt{p_2},...,\sqrt{p_n})&
T_n(\sqrt{p_1},\sqrt{p_2},...,\sqrt{p_n})&{\bf 0}\\
T_n^T(\sqrt{p_1},\sqrt{p_2},...,\sqrt{p_n})&{\bf 0}&{\bf 0}\\
{\bf 0}&{\bf 0}&1
\end{array}
\right ),
$$
$$
V_{a_2}=
\left (
\begin{array}{ccccc}
0&{\bf 0}&1&{\bf 0}&0\\
{\bf 0}&R_{n-1}(\sqrt{p_2},...,\sqrt{p_n})&
{\bf 0}&T_{n-1}(\sqrt{p_2},...,\sqrt{p_n})&{\bf 0}\\
1&{\bf 0}&0&{\bf 0}&0\\
{\bf 0}&T_{n-1}^T(\sqrt{p_2},...,\sqrt{p_n})&
{\bf 0}&{\bf 0}&{\bf 0}\\
0&{\bf 0}&0&{\bf 0}&1
\end{array}
\right ),
$$
$$...,$$
$$
V_{a_k}=
\left (
\begin{array}{ccccc}
{\bf 0}&{\bf 0}&I_{k-1}&{\bf 0}&{\bf 0}\\
{\bf 0}&R_{n+1-k}(\sqrt{p_k},...,\sqrt{p_n})&
{\bf 0}&T_{n+1-k}(\sqrt{p_k},...,\sqrt{p_n})&{\bf 0}\\
I_{k-1}&{\bf 0}&{\bf 0}&{\bf 0}&{\bf 0}\\
{\bf 0}&T_{n+1-k}^T(\sqrt{p_k},...,\sqrt{p_n})&
{\bf 0}&{\bf 0}&{\bf 0}\\
{\bf 0}&{\bf 0}&{\bf 0}&{\bf 0}&1
\end{array}
\right ),
$$
$$...,$$
$$
V_{a_n}=
\left (
\begin{array}{cccc}
{\bf 0}&{\bf 0}&I_{n-1}&{\bf 0}\\
{\bf 0}&1&{\bf 0}&{\bf 0}\\
I_{n-1}&{\bf 0}&{\bf 0}&{\bf 0}\\
{\bf 0}&{\bf 0}&{\bf 0}&1
\end{array}
\right ),
$$
$$
V_\$=
\left (
\begin{array}{cc}
{\bf 0}&I_n\\
I_n&{\bf 0}
\end{array}
\right ).
$$

{\it Case 1.} The input is $\kappa a_1^*a_2^*...a_n^*\$$.

The starting superposition is $\ket{q_1}$.
After reading the left endmarker
the superposition becomes
$\sqrt{p_1}\ket{q_1}+\sqrt{p_2}\ket{q_2}+\ldots+\sqrt{p_n}\ket{q_n}$
and after reading $a_1^*$ the superposition remains the same.

If the input contains $a_k$ then
reading the first $a_k$ changes the non-halting part of the superposition to
$\sqrt{p_k}\ket{q_k}+\ldots+\sqrt{p_{n}}\ket{q_{n}}$
and after reading all the rest of $a_k$
the non-halting part of the superposition remains the same.

Reading the right endmarker maps $\ket{q_{n}}$ to $\ket{q_{2n}}$.
Therefore, the superposition after reading it contains
$\sqrt{p_n}\ket{q_{2n}}$.
This means that the automaton accepts with probability $p_n$
because $q_{2n}$ is an accepting state.\\

{\it Case 2.} The input is $\kappa a_1^*a_2^*...a_k^*a_ka_m...\ (k>m)$.

After reading the last $a_k$ the non-halting part of the superposition is
$\sqrt{p_k}\ket{q_k}$\\ $+\ldots+\sqrt{p_{n}}\ket{q_{n}}$.
Then reading $a_m$
changes the non-halting part to\\
$\frac{\sqrt{p_m}(p_k+...+p_n)}{(p_m+...+p_n)}\ket{q_m}+\ldots+
\frac{\sqrt{p_n}(p_k+...+p_n)}{(p_m+...+p_n)}\ket{q_n}.$
This means that the automaton accepts with probability
$\leq \frac{p_n(p_k+...+p_n)^2}{(p_m+...+p_n)^2}$ and rejects
with probability at least
$$1-\frac{p_n(p_k+...+p_n)^2}{(p_m+...+p_n)^2}
\geq 1-\frac{p_n(p_k+...+p_n)^2}{(p_{k-1}+...+p_n)^2}=p_n$$
that follows from (1).
\qed

\begin{corollary}
\label{cor1}
The language $L_n$ can be recognized by a 1-way QFA with
the probability of correct answer at least $\frac{1}{2}+\frac{c}{n}$,
for a constant $c$.
\end{corollary}

\noindent
{\bf Proof:}
By resolving the equation $p^{\frac{n+1}{n-1}}+p=1$, we get
$p=\frac{1}{2}+\Theta(\frac{1}{n})$.
\qed

\begin{theorem}
\label{part2}
The language $L_n$ cannot be recognized by a 1-way QFA with probability
greater than $p$ where $p$ is the root of
\begin{equation}
\label{e1}
(2p-1)=\frac{2(1-p)}{n-1}+4\sqrt{\frac{2(1-p)}{n-1}}
\end{equation}
in the interval $[1/2, 1]$.
\end{theorem}

\noindent
{\bf Proof:}
Assume we are given a 1-way QFA $M$. 
We show that, for any $\epsilon>0$, there is a word 
such that the probability of correct answer is less than $p+\epsilon$.
 

\begin{lemma}
\cite{AF 98}
\label{LT1}
Let $x\in \Sigma^{+}$.
There are subspaces $E_1$, $E_2$ such that $E_{non}=E_1\oplus E_2$ and
\begin{enumerate}
\item[(i)]
If $\psi\in E_1$, then $V_x(\psi)\in E_1$,
\item[(ii)]
If $\psi\in E_2$, then $\| V'_{x^k}(\psi)\|\rightarrow 0$ when
$k\rightarrow\infty$.
\end{enumerate}
\end{lemma}

We use $n-1$ such decompositions: for $x=a_2$,
$x=a_3$, $\ldots$, $x=a_n$.
The subspaces $E_1$, $E_2$ corresponding to $x=a_m$ are
denoted $E_{m, 1}$ and $E_{m, 2}$.

Let $m\in\{2, \ldots, n\}$, $y\in a_1^* a_2^* \ldots a_{m-1}^*$.
Remember that $\psi_{y}$ denotes the superposition after reading $y$
(with observations w.r.t. $E_{non}\oplus E_{acc}\oplus E_{rej}$
after every step).
We express $\psi_y$ as $\psi^1_{y}+\psi^2_{y}$, $\psi^1_{y}\in E_{m, 1}$,
$\psi^2_{y}\in E_{m, 2}$.


{\em Case 1.}
$\|\psi^2_{y}\|\leq \sqrt{\frac{2(1-p)}{n-1}}$
for some $m\in\{2, \ldots, n\}$ and $y\in a_1^* \ldots a_{m-1}^*$.

Let $i>0$.
Then, $y a_{m-1}\in L_n$ but $y a_m^i a_{m-1}\notin L_n$.
Consider the distributions of probabilities on $M$'s answers ``accept"
and ``reject" on $y a_{m-1}$ and $y a_m^i a_{m-1}$.
If $M$ recognizes $L_n$ with probability $p+\epsilon$,
it must accept $y a_{m-1}$ with probability at least $p+\epsilon$
and reject it with probability at most $1-p-\epsilon$.
Also, $y a_m^i a_{m-1}$ must be rejected with
probability at least $p+\epsilon$
and accepted with probability at most $1-p-\epsilon$.
Therefore, both the probabilities of accepting and the probabilities of
rejecting must differ by at least
\[ (p+\epsilon)-(1-p-\epsilon)=2p-1+2\epsilon .\]
This means that the {\em variational distance} between two probability
distributions (the sum of these two distances) must be
at least $2(2p-1)+4\epsilon$.
We show that it cannot be so large.

First, we select an appropriate $i$.
Let $k$ be so large that $\|V'_{a_m^k}(\psi^2_{y})\|\leq \delta$ for
$\delta=\epsilon/4$.
$\psi^1_{y}, V'_{a_m}(\psi^1_{y}), V'_{a_m^2}(\psi^1_{y})$, $\ldots$
is a bounded sequence in a finite-dimensional space. Therefore, it
has a limit point and there are $i, j$ such that
\[  \|V'_{a_m^j}(\psi^1_{y})-V'_{a_m^{i+j}}(\psi^1_{y})\|<\delta.\]
We choose $i, j$ so that $i>k$.

The difference between the two probability distributions comes from two sources.
The first source is the difference between $\psi_{y}$ and
$\psi_{y a_m^i}$ (the states of $M$ before reading $a_{m-1}$).
The second source is the possibility of $M$ accepting while
reading $a_m^i$ (the only part that is different in the two words).
We bound each of them.

The difference $\psi_{y}-\psi_{y a_{m}^i}$
can be partitioned into three parts.
\begin{equation}
\label{eq5}
 \psi_{y}-\psi_{y a_{m}^i}=(\psi_{y}-\psi^1_{y})+
(\psi^1_{y}-V'_{a_m^i}(\psi^1_{y}))+
(V'_{a_m^i}(\psi^1_{y})-\psi_{y a_{m}^i}).
\end{equation}

The first part is
$\psi_{y}-\psi^1_{y}=\psi^2_{y}$ and
$\|\psi^2_{y}\|\leq\sqrt{\frac{2(1-p)}{n-1}}$.
The second and the third parts are both small.
For the second part, notice that $V'_{a_m}$ is unitary
on $E_{m, 1}$ (because $V_{a_m}$ is unitary and $V_{a_m}(\psi)$ does not
contain halting components for $\psi\in E_{m, 1}$).
Hence, $V'_{a_m}$ preserves distances on $E_{m, 1}$ and
\[ \|\psi^1_{y}-V'_{a_m^i}(\psi^1_{y})\|=
\|V'_{a_m^j}(\psi^1_{y})-V'_{a_m^{i+j}}(\psi^1_{y}) \| <\delta \]





For the third part of (\ref{eq5}), remember that 
$\psi_{y a_m^i}=V'_{a_m^i}(\psi_y)$.
Therefore,
\[ \psi_{y a_{m}^i}-V'_{a_m^i}(\psi^1_y) = V'_{a_m^i}(\psi_y) -
V'_{a_m^i}(\psi^1_y) = V'_{a_m^i}(\psi_y-\psi^1_y) = V'_{a_m^i}(\psi^2_y) \]
and $\|\psi^2_{y a_{m}^i}\|\leq \delta$ because $i>k$.
Putting all three parts together, we get
\[ \|\psi_{y}-\psi_{y a_{m}^i}\|\leq \|\psi_{y}-\psi^1_{y}\|+
\|\psi^1_{y}-\psi^1_{y a_{m}^i}\|+
\|\psi^1_{y a_{m}^i}-\psi_{y a_{m}^i}\|\leq 
\sqrt{\frac{2(1-p)}{n-1}}+ 2\delta.\]

\begin{lemma}
\cite{BV 97}
\label{BVTheorem}
Let $\psi$ and $\phi$ be such that $\|\psi\|\leq 1$, $\|\phi\|\leq 1$ and
$\|\psi-\phi\|\leq\epsilon$.
Then the total variational distance resulting from measurements
of $\phi$ and $\psi$ is at most $4\epsilon$.
\end{lemma}

This means that the difference between any probability distributions
generated by $\psi_{y}$ and $\psi_{y a^i_{m}}$ is at most 
\[ 4\sqrt{\frac{2(1-p)}{n-1}}+ 8\delta.\]
In particular, this is true for the probability distributions obtained 
by applying $V_{a_{m-1}}$, $V_{\$}$ and the corresponding measurements
to $\psi_{y}$ and $\psi_{y a_m^i}$.

The probability of $M$ halting while reading $a_m^i$ is at most
$\|\psi^2_{\cents}\|^2=\frac{2(1-p)}{n-1}$. Adding it increases the
variational distance by at most $\frac{2(1-p)}{n-1}$.
Hence, the total variational distance is at most 
\[ \frac{2(1-p)}{n-1}+4\sqrt{\frac{2(1-p)}{n-1}}+ 8\delta=
\frac{2(1-p)}{n-1}+4\sqrt{\frac{2(1-p)}{n-1}}+ 2\epsilon .\]
By definition of $p$, this is the same as $(2p-1)+2\epsilon$.
However, if $M$ distinguishes $y$ and $y a_m^i$ correctly,
the variational distance must be at least $(2p-1)+4\epsilon$.
Hence, $M$ does not recognize one of these words correctly.

{\em Case 2.} 
$\|\psi^2_{y}\|> \sqrt{\frac{2(1-p)}{n-1}}$
for every $m\in\{2, \ldots, n\}$ and $y\in a_1^* \ldots a_{m-1}^*$.

We define a sequence of words $y_1, y_2, \ldots, y_m\in a_1^* \ldots a_n^*$.
Let $y_1=a_1$ and $y_{k}=y_{k-1} a_k^{i_k}$ for $k\in\{2, \ldots, n\}$ where
$i_k$ is such that
\[ \|V'_{a_k^{i_k}}(\psi^2_{y_{k-1}})\|
\leq \sqrt{\frac{\epsilon}{n-1}}.\]
The existence of $i_k$ is guaranteed by (ii) of Lemma \ref{LT1}.

We consider the probability that $M$ halts on
$y_n=a_1 a_2^{i_2} a_3^{i_3} \ldots a_n^{i_n}$
before seeing the right endmarker.
Let $k\in\{2, \ldots, n\}$. 
The probability of $M$ halting while reading the $a_k^{i_k}$ part
of $y_n$ is at least
\[ \| \psi^2_{y_{k-1}}\|^2 -
\| V'_{a_k^{i_k}}(\psi^2_{y_{k-1}}) \|^2 
> \frac{2(1-p)}{n-1}-\frac{\epsilon}{n-1} .\]
By summing over all $k\in\{2, \ldots, n\}$, the probability
that $M$ halts on $y_n$ is at least
\[ (n-1)\left(\frac{2(1-p)}{n-1}-  \frac{\epsilon}{n-1}\right)=
2(1-p)-\epsilon .\]
This is the sum of the probability of accepting and the probability
of rejecting. Hence, one of these two probabilities must be at least
$(1-p)-\epsilon/2$. Then, the probability of the opposite 
answer on any extension of $y_n$ is at most $1-(1-p-\epsilon/2)=p+\epsilon/2$.
However, $y_n$ has both extensions that are in $L_n$ and extensions that are not.
Hence, one of them is not recognized with probability $p+\epsilon$.
\qed

By solving the equation (\ref{e1}), we get

\begin{corollary}
\label{cor2}
$L_n$ cannot be recognized with probability greater than
$\frac{1}{2}+\frac{3}{\sqrt{n-1}}$.
\end{corollary}

\noindent
{\bf Proof:}
The right-hand side of (\ref{e1}) is at most 
$\frac{1}{n-1}+4\sqrt{\frac{1}{n-1}}$ because $p\geq 1/2$ and,
hence, $1-p\leq 1/2$. This implies
\[ 2p-1 \leq \frac{1}{n-1}+4\sqrt{\frac{1}{n-1}}, \]
\[ p\leq \frac{1}{2}+2\sqrt{\frac{1}{n-1}}+\frac{1}{2(n-1)}\leq
\frac{1}{2}+3\sqrt{\frac{1}{n-1}} \]
and $L_n$ cannot be recognized with probability greater than $p$
by Theorem \ref{part2}.
\qed

Let $n_1=2$ and $n_k=\frac{9n_{k-1}^2}{c^2}+1$ for $k>1$
(where $c$ is the constant from Theorem \ref{part1}).
Also, define $p_k=\frac{1}{2}+\frac{c}{n_k}$.
Then, Corollaries \ref{cor1} and \ref{cor2} imply

\begin{theorem}
For every $k>1$,
$L_{n_k}$ can be recognized with by a 1-way QFA with the probability 
of correct answer $p_{k}$ but cannot be recognized with the 
probability of correct answer $p_{k-1}$.
\end{theorem}

\noindent
{\bf Proof:}
By Corollary \ref{cor1}, $L_{n_k}$  can be recognized with 
probability $\frac{1}{2}+\frac{c}{n_k}=p_k$.

On the other hand, by Corollary \ref{cor2}, $L_{n_k}$ cannot be
recognized with probability $\frac{1}{2}+\frac{3}{\sqrt{n_k-1}}$.
The definition of $n_k$ implies 
$n_k-1=\frac{9n_{k-1}^2}{c^2}$, $\sqrt{n_k-1}=\frac{3 n_{k-1}}{c}$,
\[ \frac{1}{2}+\frac{3}{\sqrt{n_k-1}}=\frac{1}{2}+\frac{c}{n_{k-1}}=p_{k-1} .\]
\qed

Thus, we have constructed a sequence of languages $L_{n_1}$, $L_{n_2}$, $\ldots$
such that, for each $L_{n_k}$, the probability with which $L_{n_k}$ 
can be recognized by a 1-way QFA is smaller than for $L_{n_{k-1}}$.

Our final theorem is a counterpart of Theorem 
\ref{T2}. It generalizes Theorem \ref{T4}.

\begin{theorem}
\label{T9}
Let $L$ be a language and $M$ be its minimal automaton.
If there is no $q_1, q_2, q_3, x, y$ such that
\begin{enumerate}
\item~
the states $q_1 , q_2 , q_3$ are pairwise different,
\item
If $M$ starts in the state $q_1$ and reads $x$,
it passes to $q_2$,
\item
If $M$ starts in the state $q_2$ and reads $x$,
it passes to $q_2$, and
\item
If $M$ starts in the state $q_2$ and reads $y$,
it passes to $q_3$,
\item
If $M$ starts in the state $q_3$ and reads $y$,
it passes to $q_3$,
\item
both $q_2$ and $q_3$ are neither "all-accepting" state, nor
"all-rejecting" state,
\end{enumerate}
then $L$ can be recognized by
a 1-way quantum finite automaton with probability $p=0.68... $.
\end{theorem}



\end{document}